\documentclass[twocolumn,showpacs,superscriptaddress,10pt]{revtex4-2}
\usepackage{amsmath,amssymb,graphicx, subfigure, hyperref, braket, float}
\usepackage{blindtext}

\counterwithout{figure}{section}
\usepackage[export]{adjustbox}
\graphicspath{{figure/}}
\usepackage{graphicx, xcolor}
\usepackage[margin=.8in,text={10in,10.5in},centering]{geometry}
\usepackage{graphicx}
\usepackage{hyperref}
\usepackage{lipsum}
\usepackage{caption}
\usepackage{tikz}
\usepackage{needspace}  

\begin{document}
	
\title{Tunable Entangling and Steering of Ferrimagnetic Magnons via an OptoMagnoMechanical Ring}

\author{Ziyad Imara} \email{imara.ziyad@etu.uae.ac.com}
\address{Laboratory of R\&D in Engineering Sciences, Faculty of Sciences and Techniques Al-Hoceima, Abdelmalek Essaadi University, Tetouan, Morocco}

\author{Isaac Pérez Castillo} \email{iperez@izt.uam.mx}
\address{Departamento de Física, Universidad Autónoma Metropolitana-Iztapalapa, San Rafael Atlixco 186, Ciudad de México 09340, Mexico}

\author{Khadija El Anouz}\email{kelanouz@uae.ac.ma}
\address{Laboratory of R\&D in Engineering Sciences, Faculty of Sciences and Techniques Al-Hoceima, Abdelmalek Essaadi University, Tetouan, Morocco}

\author{Abderahim El Allati}\email{eabderrahim@uae.ac.ma}
\address{Laboratory of R\&D in Engineering Sciences, Faculty of Sciences and Techniques Al-Hoceima, Abdelmalek Essaadi University, Tetouan, Morocco}

\begin{abstract}
Recently, magnomechanical systems have emerged as {promising platforms} for quantum technologies, exploiting {magnon–photon–phonon} interactions to store high-fidelity quantum information.
In this paper, we propose a scheme to entangle {two spatially separated ferrimagnetic YIG crystals} by injecting a microwave field into an {optomagnonic} ring cavity.
The proposed optomagnomechanical configuration {utilizes} the coupling between {magnetostriction-induced mechanical displacements} and the optical cavity via radiation pressure. {Magnons—collective spin excitations in macroscopic ferromagnets—are} directly driven by an electromagnetic field. We {demonstrate the generation of macroscopic magnon entanglement} by exciting the optical cavity with a red-detuned microwave field and the {YIG crystals} with a blue-detuned field.
{Our analysis reveals that magnon entanglement} vanishes for identical magnomechanical couplings and {remains robust against thermal fluctuations}.
The magnon modes entangled in two ferrimagnetic crystals represent {genuine} macroscopic quantum states with potential applications in the {study} of macroscopic quantum mechanics and quantum information processing based on magnonics.
The configuration is based on {experimentally accessible parameters, providing a feasible route} of quantum technologies.
\end{abstract}

\vspace{2cm}
\maketitle
\section{Introduction}
\label{sec1}
Quantum cavity electrodynamics (QCED) describes coherent matter–electromagnetic field interactions in resonant structures \cite{A1}, providing a platform for cavity optomechanics (OM) experiments that study electromagnetic–mechanical interactions via radiation pressure forces \cite{A2,A3}. Furthermore, using {QCED}, the magnon mode in ferromagnetic materials can achieve strong coupling with microwave photons in a high-quality cavity \cite{A4,A5}, giving rise to magnon–cavity polaritons \cite{A6} through the collective excitations of a large number of spins \cite{A7}. These advances, similar to {those in} OM, gave rise to the field of magnomechanics (MM), which exploits magnetic materials in various structures \cite{A8,A9}, such as {yttrium iron} garnet (YIG) crystal in a sphere \cite{B1}, thin film \cite{B01}, bridge structure \cite{B2}, or other {crystals such as magnetic beams (e.g., CoFeB)} \cite{B3}, where magnons couple to phonons via magnetostrictive (magnetoelastic) forces. Magnon–photon and magnon–phonon couplings have made it possible to explore macroscopic quantum phenomena, such as {magnon-induced nonreciprocity} based on the magnon Kerr effect \cite{B4}, ground state cooling of {MM resonators} \cite{B5}, quantum effects such as magnon–photon–phonon {tripartite} entanglement \cite{B1,B02,B03,B04}, {indirect entanglement via magnon–magnon coupling between two squeezed magnons} \cite{B6}, as well as magnon blockade \cite{B7}.\\

Furthermore, YIG materials' recent advances in magneto-optical technology \cite{B8}, including quantum information \cite{B9}, quantum correlations \cite{B6}, and quantum networks \cite{C1}, are characterized by high-fidelity information storage via photon–magnon–phonon coupling in a hybrid MM system \cite{C2,C3}. Following the success of these two systems (i.e., OM and MM) in studying and observing numerous quantum effects {(QEs)}, an innovative optomagnomechanical (OMM) cavity system has recently emerged, where {radiation pressure enables the coupling of magnetostriction-induced mechanical displacement to the optical field} \cite{A8}. This breakthrough opens up fresh prospects for investigating enhanced nonlinear interactions, developing advanced protocols, and expanding applications in various areas of quantum technology. Moreover, theoretical research has focused on quantum entanglement in these OMM systems, including {entanglement between ferrimagnetic magnons and atomic ensembles via OMM} \cite{X2} and the entanglement of magnons with Bose–Einstein condensates \cite{C4}.\\

In hybrid MM systems, several studies have demonstrated quantum entanglement between modes in YIG spheres. J. Li and S.Y. Zhu in \cite{B02} entangled two magnonic modes via a nonlinear magnetostrictive interaction in a single cavity system. J. Li and S. Gröblacher \cite{B03} have extended this concept to two separate cavities, using a squeezed microwave field to entangle magnonic modes by quantum correlation transfer. In the same context, our investigation is inspired by the ring configuration initially proposed by S. Huang and G. S. Agarwal \cite{D1}, and M. Pinard \textit{et al.} \cite{D2} in an OM entanglement context with two moving mirrors. In this paper, we provide a scheme to prepare {macroscopic entanglement} between ferrimagnetic magnons in a novel ring–optomagnomechanical system, which consists of a dual MM component. In our proposed model, these magnons in the resonators are entangled via the nonlinear interactions of magnetostriction and radiation pressure. The low-frequency phonon mode (several tens to several hundreds of megahertz) is dispersively coupled to the magnon mode (on the order of gigahertz) via magnetostrictive interaction and to the optical cavity via radiation pressure interaction. We show this entanglement by exciting the cavity with a red-detuned microwave field and the YIG magnons with a blue-detuned microwave field. The phonon modes act as intermediaries to establish entanglement between the magnons and optical photons. The MM parametric down-conversion generates the entanglement between magnons and phonons, while the OM beam-splitter interaction allows the state-swapping between photons and phonons, yielding a macroscopic entangled state between magnons and photons.\\

The structure of the paper is as follows. A suggested OMM ring cavity model is shown in Sec.~\ref{sec2}. The quantum Langevin equations are determined in Sec.~\ref{sec3}, and the classical equations and their quantum linearization are investigated in Sec.~\ref{sec4}. In Sec.~\ref{sec5}, numerical simulations based on experimentally {accessible} parameters are used to achieve {optomagnetic quantum effects, such as entanglement and quantum steering, within the model}. The numerical results are then discussed. Lastly, Sec.~\ref{sec6} provides a conclusion of the current study.

\section{The Model Description}
\label{sec2}
\begin{figure}[H]
	\centering
	\subfigure{\includegraphics[scale=0.43]{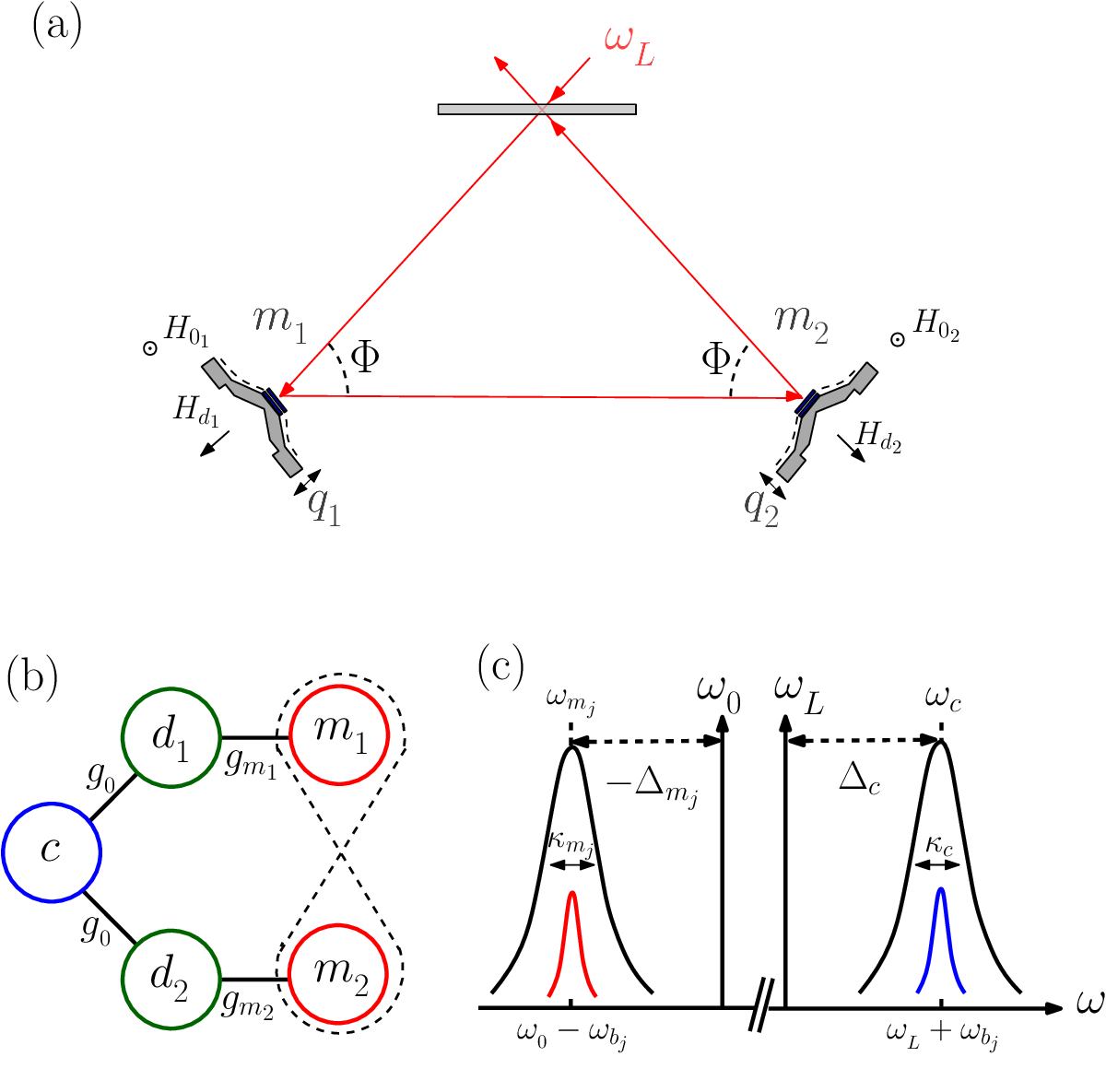}}
\caption{{\small {{(a)-(b) Schematic of the proposed opto-magnomechanical system.}} The phonon modes $d_{(1,2)}$ couple to the magnon modes $m_{(1,2)}$ by {{dispersive magnetostrictive interaction}}, and are simultaneously coupled to the optical cavity mode $c$ by {{a radiation pressure interaction}}. (c) {{System linewidths and frequencies}}. Once the cavity mode $c$ reaches resonance with the anti-Stokes (blue) sideband of the laser field at $\omega_{L}-\omega_{b_{j}}$, and the magnon mode $m_j$ reaches resonance with the Stokes (red) sideband of the microwave drive field at $\omega_{0}+\omega_{b_{j}}$, a steady entangled state between magnons and optical photons is achieved.}}
	\label{fig:1}
\end{figure}

Consider a proposed OMM cavity system, as shown in Fig.\ref{fig:1}, which consists of a {{magnon mode (Kittel mode)}} \cite{D4} in YIG resonators and an optical cavity. In the MM cavities, the mechanical vibrations are coupled to the magnons via the {{micron-sized bridge deformation of the YIG resonator}} \cite{B2}. The OM cavity consists of a fixed mirror and a {{micron-sized, highly reflective mirror attached to the YIG microbridge}} \cite{D5,D6}. The latter is {{sufficiently small and light to maintain the mechanical properties of the YIG}} \cite{D3,D7}. Furthermore, to {{model the system, we assume uniform deformation perpendicular to the fixed surfaces, with negligible bending-induced displacement}}. So with this method, one can {{closely integrate the YIG bridges with the mirrors}}. As a result, the YIG bridge and mirror can be considered a single unit, which can {{approximately oscillate at the same frequency}} \cite{A8,D8,D9}. To achieve the {{dispersive MM-optical cavity coupling}}, one can consider the {{membrane-in-the-middle approach, placing the bridge at the center of the optical cavity}} \cite{E1}. The {{magnetostrictive force is induced by aligning each YIG bridge with the biased magnetic field and amplified by driving the magnon mode with a microwave field at frequency $\omega_0$}}. The optical cavity mode is coupled to the vibration modes by {{exciting the system with an external laser at frequency $\omega_L$ through radiation pressure}}.

The Hamiltonian of the model is given as follows (we set $\hbar=1$)  
\begin{eqnarray}
	\label{1}
		\mathbf{H}&=&\omega_c c^\dagger c +\sum_{j=1,2}\omega_{m_j} m_j^\dagger m_j+\frac{\omega_{b_j}}{2}(q_{_j}^2+p_{_j}^2)\nonumber\\
		&+&\sum_{j=1,2}  \bar{g}_{0j}c^\dagger c q_{_j}+g_{m_j}m_j^\dagger m_jq_{_j}\nonumber
		\\&+&i\big(\big(Ec^\dagger e^{-i\omega_{L}t}+\Omega_j m_j^\dagger e^{-i\omega_{0}t}\big)-\text{H.C}\big),
\end{eqnarray}
where $c (c^\dagger)$ and $m_j(m_j^\dagger)$ are the annihilation (creation) operators of the optical cavity mode and the $j^{th}$ magnon (magnon$_j$) mode, respectively. Furthermore, $\mathcal{O}\equiv a,m$ satisfies $[\mathcal{O},\mathcal{O}^\dagger]=\mathbf{1}$. For {the mechanical vibration modes}, $q_j=\frac{1}{\sqrt{2}}(d_j^\dagger+d_j)$ and $p_j=\frac{i}{\sqrt{2}}(d_j^\dagger-d_j)$ are the dimensionless quadratures of position and momentum, where $d_j$ ($d_j^\dagger$) are the annihilation (creation) operators of {{the $j^{\text{th}}$ phonon mode}}, and $[q_j,p_k]=i\delta_{jk}$ is satisfied.  
$\omega_{c}$ ($\omega_{b_j}$) represents the frequency of the optical cavity ({{the $j^{\text{th}}$ phonon}}) mode.  
Moreover, to excite and generate the magnon mode, both YIG microbridges are usually placed in a uniformly biased magnetic field $H_{_0}$, and a microwave field with its magnetic component $H_{_d}$ {{is applied perpendicular to the polarization field}}.  
In addition, $\omega_{m_j}$ is the frequency of the {{$j^{\text{th}}$ magnon mode}}, which is determined by {{the external bias magnetic field}} and the gyromagnetic ratio $\gamma$ via $\omega_{m_j}=\gamma H_{_0}$, where $\gamma= 28 $GHz/T \cite{B1}.  
$\bar{g}_{_0j}$ ($ g_{m_{j}}$) is the coupling strength of the cavity mode ({{the $j^{\text{th}}$ magnon mode}}) and the {{$j^{\text{th}}$ phonon mode}}, where $\bar{g}_{_0j}=-(-1)^{{j}}{g}_{_0}\cos^2(\frac{\Phi}{2})$, and $\Phi$ is the angle between the incident light and the light reflected by the surfaces of the YIG microbridge (or, more precisely, {{by the attached mirror}}).  
Moreover, $\Omega_j=\frac{\sqrt{5}}{4}\gamma\sqrt{N_0}H_{d_j}$ denotes the Rabi frequency associated with {{the microwave drive}}, where $N_0=\rho V$ is the total number of spins, while the spin density $\rho=4.22\times10^{27}m^{-3}$ and the volume $V$ of the YIG-bridge is $5\times2 \times 1 \mu m^3$ (approximated as a cuboid), and $E=\sqrt{\frac{\kappa_{c} \mathcal{P}_L}{\hbar\omega_L}}$ is the coupling strength between the cavity and the laser, with $\mathcal{P}_L$ the power of the laser and $\kappa_{c}$ the decay rate of the optical cavity mode.

\section{Quantum Langevin equation}
\label{sec3}
Let's include {{dissipative and input noise terms}} for each mode {{under the rotating wave approximation, using the respective drive frequencies}} $\omega_L$ and $\omega_{0}$. Then, we obtain a set of quantum Langevin equations (QLEs) for the system as follows:
\begin{eqnarray}
	\label{2}
	\dot{c}&=& -i{\Delta}_{c}c - {\kappa}_{c}c - i\sum_{j=1,2}{{\bar{g}_{0j}c q_{_j}}} + E + \sqrt{2\kappa_{c}} c^{in},\nonumber\\
	\dot{m}_j &=& -i{\Delta}_{m_j}m_j - \kappa_{m_j}m_j - i g_{m_j} m_j q_j + \Omega_j + \sqrt{2\kappa_{m_j}} m^{in}_j,\nonumber\\
	\dot{q}_j&=&\omega_{b_j}p_j,\nonumber\\
	\dot{p}_j&=& -\omega_{b_j}q_j - g_{_{mj}} m_j^\dagger m_j - \bar{g}_{0j} c^\dagger c - \gamma_{b_j} p_j + \xi_j,
\end{eqnarray}
where ${\Delta}_{c} = \omega_{c} - \omega_L$, ${\Delta}_{m_j} = \omega_{m_j} - \omega_0$.  
$\kappa_m$ is the decay rate of the magnon mode, and $\gamma_{b_j}$ is the damping rate of the {{$j^{\text{th}}$ phonon mode}}.  
Moreover, $\varphi^{in}$ is the $\varphi$-mode input noise operator which has zero mean \cite{E2} and is {{characterized}} by the following correlation functions:
\begin{equation}
	\braket{\varphi^{in\dagger}(t)\varphi^{in}(t') ; \varphi^{in}(t)\varphi^{in\dagger}(t')} = (\bar{n}_\varphi ; \bar{n}_\varphi + 1) \delta(t - t'),
\end{equation}
with $\bar{n}_\varphi = \left(\exp\left(\frac{\hbar\omega_\varphi}{k_B T}\right) + 1\right)^{-1}$ {{denoting the mean thermal occupation number of the $\varphi$-mode ($\varphi = c, m_j$)}} at bath temperature $T$ and the Boltzmann constant $k_B$ \cite{E3}.  
Moreover, $\xi_j$ is the Brownian noise operator of the{{$j^{\text{th}}$ mechanical mode}} with zero mean value $\braket{\xi_j} = 0$. In the Markov approximation, this correlation function for a {{mechanical mode with high quality factor $\mathcal{Q}_j = \omega_{b_j} / \gamma_{b_j} \gg 1$ takes the form}} \cite{E4}
\begin{equation}
	\label{3}
	\braket{\xi_j(t)\xi_k(t') + \xi_j(t')\xi_k(t)} / 2 = \gamma_{b_j} (2\bar{n}_{b_j} + 1) \delta_{jk} \delta(t - t'),
\end{equation}
where $\bar{n}_{b_j} = \left(\exp\left(\frac{\hbar\omega_{b_j}}{k_B T}\right) + 1\right)^{-1}$.

\section{Classical and linearized quantum dynamics}
\label{sec4}
Using {{strong coherent driving fields on the optical cavity and magnon modes leads to large steady-state amplitudes}} $|\alpha_{s}|$,$|m_{js}|\gg1$, {{thus requiring the QLEs in Eq.~(\ref{2}) to be linearized around the steady-state value $\mathcal{O}_s$ as $\mathcal{L} = \mathcal{L}_s + \tilde{\mathcal{L}}$, where $\mathcal{L} = c, m_j, q_j, p_j$}}.\\

As a result, we can obtain a set of algebraic equations for the steady-state values as follows:
\begin{eqnarray}
	\label{4}
	\alpha_{s}&= \braket{c}=&\frac{E}{i\tilde{\Delta}_{c}+\kappa_{c}},~~	m_{js}=\braket{m_j}=\frac{\Omega_{j}}{i\tilde{\Delta}_{m_j}+\kappa_{m_j}},\nonumber\\ 
q_{js}&=& -\frac{1}{\omega_{b_{j}}}(\bar{g}_0|\alpha_{s}|^2+{g}_{m_j}|m_{js}|^2),
\end{eqnarray}
where $\tilde{\Delta}_{c}=\Delta_{c}+\sum_{j=1,2}\bar{g}_{_0j}q_{js}$ and $\tilde{\Delta}_{m_j}={\Delta}_{m_j}+g_{_{mj}}{q_{js}}$ are the effective detuning of the cavity drive and {{the $j^{\text{th}}$ magnon drive}}, respectively.  
These {{detunings include frequency shifts induced by mechanical displacement $q_{js}$ in both phonon modes}}.\\

The linearized QLEs describe the quantum fluctuations of the system using the quadrature fluctuation operators
\begin{equation}
 \Xi(t)=  [\sqrt{2{\kappa}_{c}}\tilde{\Theta}_{c},\Xi_{m_1},\Xi_{m_2}],
\end{equation}
 where $\Xi_{m_j}= \sqrt{2{\kappa}_{m_j}}\tilde{\Theta}_{m_j}\oplus[\tilde{q}_j,\tilde{p}_j]$,  
$\tilde{\Theta}_\varphi=[\tilde{I}_\varphi,\tilde{J}_\varphi]$ and $\tilde{I}_\varphi={ (\tilde{\varphi} +\tilde{\varphi}^\dagger)/}{\sqrt{2}}$, $\tilde{J}_\varphi=i{(\tilde{\varphi}^\dagger-\tilde{\varphi})/}{\sqrt{2}}$, with ($\varphi=c,m_j$). Moreover, we also {{neglect second-order fluctuation terms, assuming mean values dominate over quantum fluctuations}} (see \textbf{Appendix A} for more details); it can be written in matrix form by the following formula:
\begin{equation}
	\label{5}
	\dot{\Xi}(t)=\Upsilon\Xi(t)+\mathcal{N}(t),
\end{equation}
with $\Upsilon$ as the drift matrix, given as 
\begin{equation}
	\label{6}
	{\Upsilon=\begin{pmatrix}
			\Upsilon_c & \Upsilon_0' & -\Upsilon_0'\\
			\Upsilon_0 & \Upsilon_{m_1} &\mathbf{0}_{_{4,4}}\\
			-\Upsilon_0 & \mathbf{0}_{_{4,4}} & \Upsilon_{m_2} &\\
	\end{pmatrix}},
\end{equation}
and 
\begin{equation}
	\label{7}
	{ \Upsilon_c=\begin{pmatrix}
		-\kappa_{c} & \tilde{\Delta}_{c}\\
		-\tilde{\Delta}_{c}& -\kappa_{c} \\
	\end{pmatrix}},
\end{equation}
 
\begin{equation}
	\label{8}
	\Upsilon_{m_j} =\begin{pmatrix}
			-\kappa_{m_j} & \tilde{\Delta}_{m_j} & -G_{m_j} & 0  \\
			-\tilde{\Delta}_{m_j} & -\kappa_{m_j} & 0 & 0  \\
			0 & 0 & 0 & \omega_{b_j}  \\
			0 & G_{m_j} & -\omega_{b_{j}} & -\gamma_{b_j}  \\
	\end{pmatrix},
\end{equation}
where $G_{_{0}}=i\sqrt{2}{g}_{_{0}}\alpha_{s}\cos^2(\Phi/2)=i\tilde{G}_{_{0}}\cos^2(\Phi/2)$, $G_{mj}=i\sqrt{2}g_{mj}m_{js}$ are the effective opto and magnomechanical coupling forces, respectively. $\mathbf{0}_{_{l,c}}$ is the zero matrix with $l$: rows and $c$: columns, and 
\begin{equation}
	\label{9}
	\Upsilon_{0}^T =\begin{pmatrix}
		0 & 0 & 0 & G_{_0} \\
		0 & 0 & 0 & 0  \\
	\end{pmatrix},\\
	\Upsilon_0' =\begin{pmatrix}
		0 & 0 & -G_0 & 0  \\
		0 & 0 & 0 & 0  \\
	\end{pmatrix}.
\end{equation}
$\mathcal{N}(t)$ denotes the noise vector of the form $\mathcal{N}(t)=[\sqrt{2{\kappa}_{c}}\tilde{\Theta}^{in}_{c},\mathcal{N}_{m_1},\mathcal{N}_{m_2}]^T$, where $\mathcal{N}_{m_j}=\sqrt{2{\kappa}_{m_j}}\tilde{\Theta}^{in}_{m_j}\oplus[0,\xi_j]$, and $\tilde{\Theta}^{in}_\varphi=[\tilde{I}^{in}_\varphi,\tilde{J}^{in}_\varphi]$. $\tilde{I}^{in}_{\varphi}$ and $\tilde{J}^{in}_\varphi$ are the quadrature {{components of the input noise for the $\varphi$-mode}}.

The drift matrix in Eq.~(\ref{5}) is given under the condition $|\tilde{\Delta}_{c}|,|\tilde{\Delta}_{m_j}|\simeq\omega_{b_j}\gg \kappa_{c},\kappa_{m_j}$, which is the optimal condition for quantum correlations in the system \cite{B1}, resulting in the following simplified approximate expressions $\alpha_{s}\simeq{E}/{i\tilde{\Delta}_{c}}$ and $m_{js}\simeq\Omega_j/i\tilde{\Delta}_{m_j}$, which are {{approximately imaginary, resulting in real effective couplings}} $G_{0}$ and $G_{mj}$.\\

As a result of the linearized dynamics and the Gaussian nature of the quantum noise, its state can be fully described as the steady state of the quantum fluctuations of the system, {{characterized by the covariance matrix $\mathcal{V}$ with components}}  
\begin{equation}
\mathcal{V}_{ij}(t)=\frac{1}{2}\braket{\Xi_{i}(t)\Xi_{j}(t')+\Xi_{j}(t')\Xi_{i}(t)}.
\end{equation}
Hence, one can obtain the steady state $\mathcal{V}$ by solving the Lyapunov equation as follows:
\begin{equation}
	\label{10}
	\Upsilon\mathcal{V}+\mathcal{V}\Upsilon^T=-\beth,
\end{equation}
where $\beth$ is the diffusion matrix defined by
\begin{equation}
	\label{11}
	\beth{ij}\delta(t'-t)=\frac{1}{2}\braket{n_i(t)n_j(t')+n_j(t')n_i(t)}. 
\end{equation}
It can be rewritten as 
\begin{equation}
\beth=[\beth_c\oplus\beth_{m_1}\oplus\beth_{m_2}],
\end{equation}
where
\begin{eqnarray}
 \beth_{c}&=&\text{diag} [{\kappa}_{c}\bar{N}_{c},\kappa_{c}\bar{N}_{c}],\nonumber\\
  \beth_{m_j}&=&\text{diag} [\kappa_{m_j}\bar{N}_{m_j},\kappa_{m_j}\bar{N}_{m_j},0,\gamma_{b_j}\bar{N}_{b_j}],
\end{eqnarray} 
with $\bar{N}_\varphi=2\bar{n}_{\varphi}+1$, and $j=(1,2)$.

\section{Quantum entanglement and Quantum Steering}
\label{sec5}
In this section, we investigate the quantum entanglement of the {{two-mode Gaussian state using the covariance matrix $\mathcal{V}$ from Eq.~\ref{10}}}. To quantify bipartite entanglement, we employ the logarithmic negativity $\mathcal{E}_N$ \cite{E5,E6}. Next, we {{note that these measures between subsystems $\mathcal{A}$ and $\mathcal{B}$ are computed from the $4\times 4$ reduced covariance matrix $\mathcal{C_V}_{R}$ of $\mathcal{V}$}}. The reduced matrix $\mathcal{C_V}_{R}$ takes the following form
\begin{equation}
	\label{12}
	\mathcal{C_V}_{R}=\begin{pmatrix}
		\mathcal{A}&\mathcal{C}\\
		\mathcal{C}^T&\mathcal{B}
	\end{pmatrix}, 
\end{equation}
$\mathcal{A}_{(2 \times 2)}$, $\mathcal{B}_{(2 \times 2)}$ and $\mathcal{C}_{(2 \times 2)}$ are sub-block matrices of $\mathcal{C_V}_{R}$. The logarithmic negativity $\mathcal{E}_N$ is defined as: 
\begin{equation}
	\label{13}
	\mathcal{E}_N=\max\big\{0,-\ln(2 \eta)\big\},
\end{equation}
where 
\begin{equation}
	\label{14}
	\eta= 2^{-1/2}\big[\chi(\mathcal{C_V}_{R})-\sqrt{(\chi(\mathcal{C_V}_{R})^2-4 \det(\mathcal{C_V}_{R}))}\big]^{1/2},
\end{equation}
with $\chi(\mathcal{C_V}_{R})=\alpha + \beta - 2\gamma$, {{where $\alpha = \det(\mathcal{A})$, $\beta = \det(\mathcal{B})$, and $\gamma = \det(\mathcal{C})$}}.\\

Here, we introduce the notion of a quantum steering criterion based on the quantum coherent information framework {{as recently proposed by}} \cite{C4,Y1}, for arbitrary bipartite Gaussian states $\mathcal{A}$ and $\mathcal{B}$. This approach utilizes $\mathcal{C_V}_{R}$ {{analogously to the entanglement computation above}}. The Gaussian steering $\mathcal{A}\to\mathcal{B}$ (respectively, from $\mathcal{B}\to\mathcal{A}$) can be quantified using the following formula:
\begin{eqnarray}
	\mathcal{S}^{_{\mathcal{A}\to\mathcal{B}}}&=&\text{max}\Big\{0,\frac{1}{2}\ln(\frac{\alpha}{4\delta})\Big\}, \nonumber\\
	\mathcal{S}^{_{\mathcal{B}\to\mathcal{A}}}&=&\text{max}\Big\{0,\frac{1}{2}\ln(\frac{\beta}{4\delta})\Big\}.
\end{eqnarray}
The {{steering asymmetry can then be quantified using the definitions above}}:
\begin{equation}\label{AS}
	\mathcal{S}_{\mathcal{N}}=|\mathcal{S}^{_{\mathcal{A}\to\mathcal{B}}}-\mathcal{S}^{_{\mathcal{B}\to\mathcal{A}}}|.
\end{equation} 
{{which characterizes the asymmetry in the steering behavior of the two-mode Gaussian state}}.\\

\begin{figure*}[ht]
	\centering
	\begin{minipage}{\textwidth}
		\centering
		\subfigure{\label{A4}\includegraphics[scale=0.308]{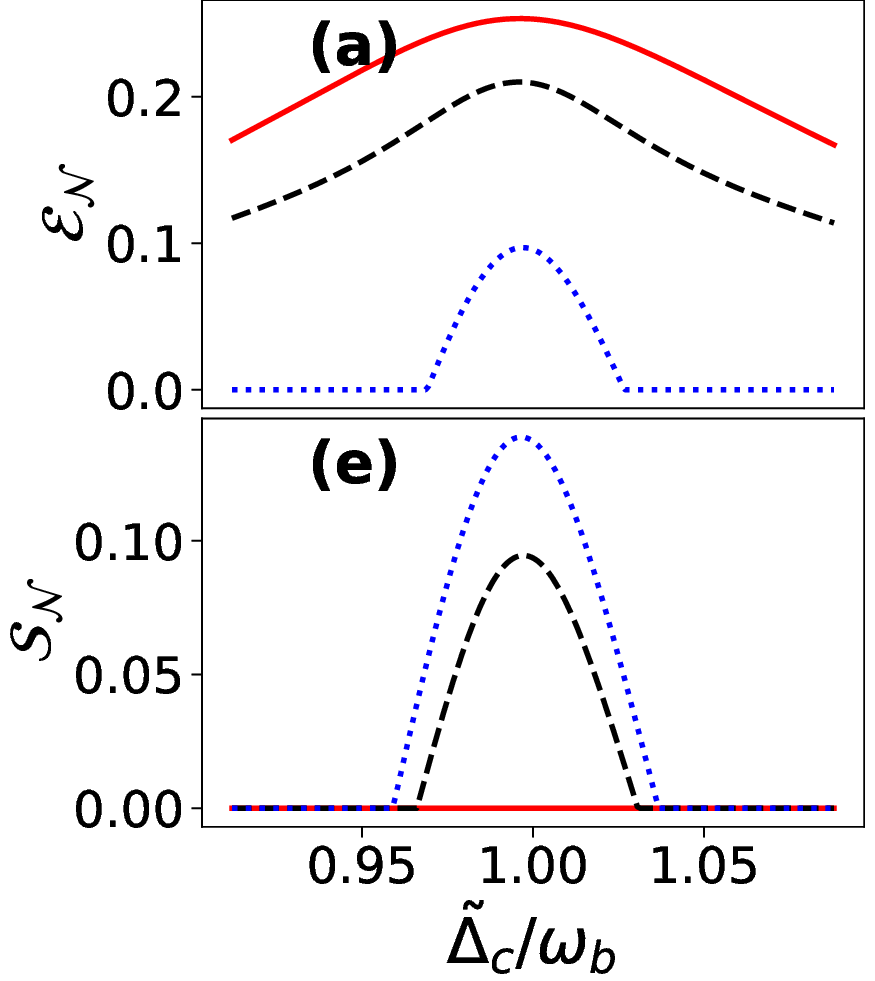}}
		\hfill
		\subfigure{\label{A6}\includegraphics[scale=0.302]{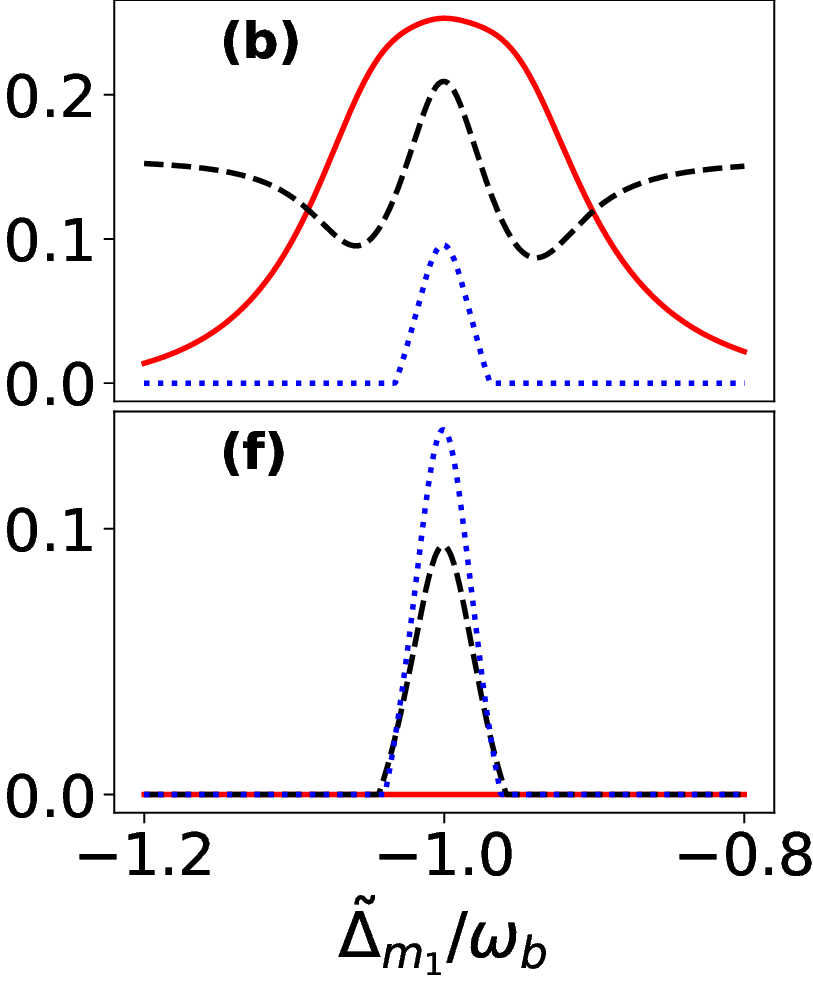}}
		\hfill
		\subfigure{\label{A4}\includegraphics[scale=0.302]{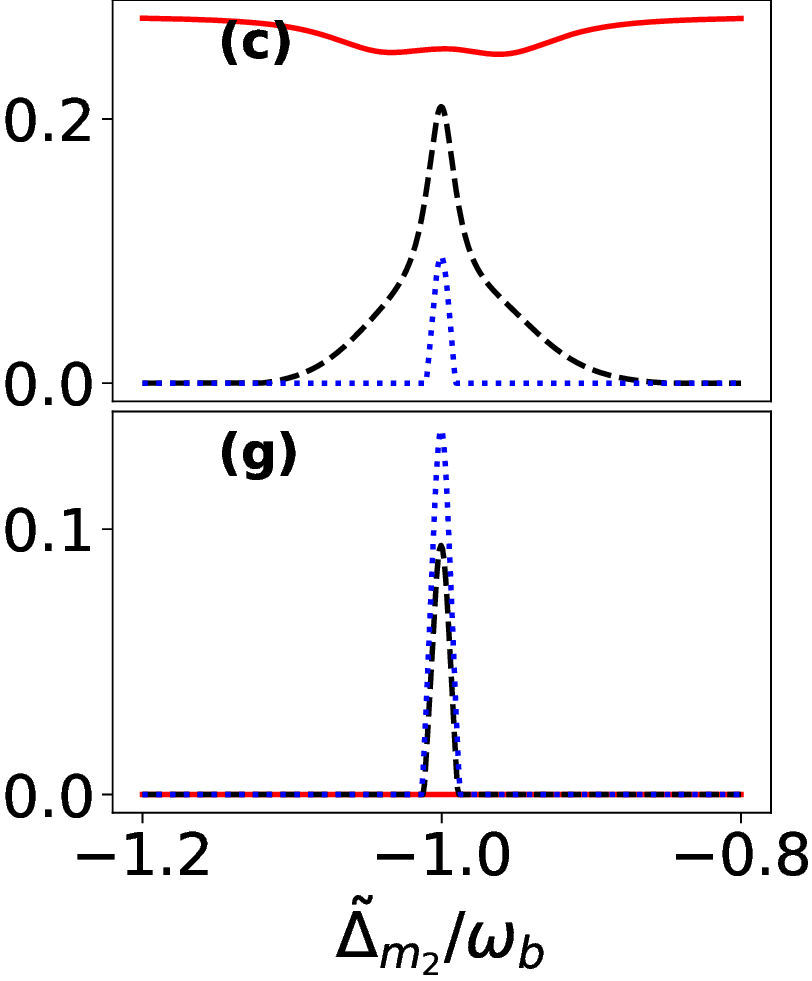}}
		\hfill
		\subfigure{\label{A6}\includegraphics[scale=0.302]{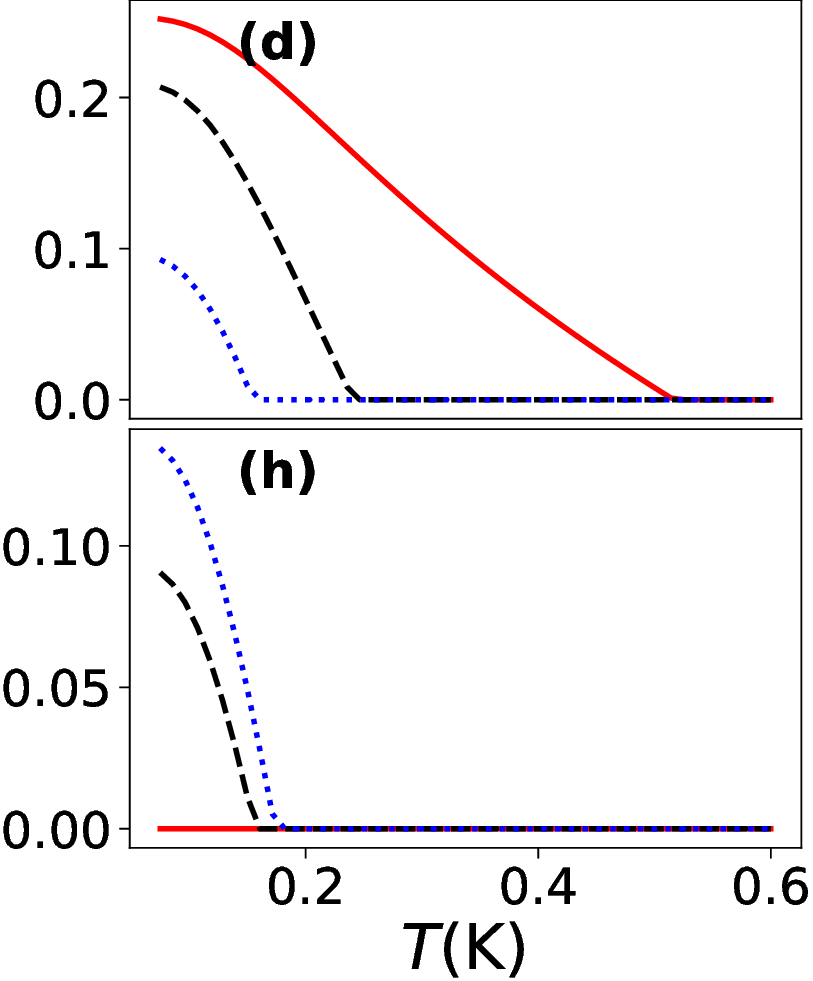}}
	\end{minipage}
   \captionsetup{justification=raggedright, singlelinecheck=false}
	\caption{{{\small  Optical cavity-magnon$_1$ (solid), optical cavity-magnon$_2$ (dashed), and magnon$_1$-magnon$_2$ (doted) {{in (a)-(d)}} entanglement and {{in (e)-(h)}} steering versus (a,e) the effective parameter of the detuned cavity $\tilde{\Delta}_c$, (b,f) the effective parameter of the detuned magnon$_1$ $\tilde{\Delta}_{m_1}$, (c,g) $\tilde{\Delta}_{m_2}$, and (d,h) bath temperature. We take $\tilde{\Delta}_c=-\tilde{\Delta}_{m_{(1,2)}}\simeq\omega_{b}$, $\omega_{b_{(1,2)}}=\omega_{b}$ and {{$T = 10$ mK}}. Please refer to the text for more information on the other parameters.}}}
	\label{fig:2}
\end{figure*}

Note that the frequency of the mechanical mode is the lowest in the system and is expressed in {{megahertz}}. Thus, although at the lowest temperature achievable by a dilution refrigerator, this mode remains {{highly thermally populated}}. Therefore, it must be cooled to a temperature close to its ground state to allow the preparation of quantum states of the system \cite{E7}. To achieve this cooling in the low-frequency mechanical {{mode}}, the optical cavity was driven using a red-detuned microwave field at frequency $\omega_{c}-\omega_{b}$, corresponding to resonance with the OM anti-Stokes sideband.  
The magnon modes are driven by a blue-detuned microwave field $\omega_{m_j}+\omega_{b}$, {{which is resonant with}} the MM Stokes sideband. Under the condition $\omega_{b}\gg G,\kappa$, which ensures that the system operates in the resolved sideband regime {{—}} achievable for a micron-sized YIG bridge \cite{B2} coupled to a typical OM cavity \cite{A1} {{—and justifies}} the rotating wave approximation, the dispersive OM and MM couplings enable parametric down-conversion (PDC) to be achieved for a blue-detuned drive and state-swap (beam-splitter) operation for a red-detuned drive. The phonon modes are used to achieve entanglement between the magnon and optical modes. At the same time, the MM PDC, which leads to the entanglement of magnons and phonons, as well as the OM beam-splitter interaction, which allows {{state swapping}} between photons and phonons, probes the entanglement between magnons and photons. In conclusion, it can be affirmed that the strong red-detuned microwave drive works in two senses, cooling the mechanical mode and probing the MM entanglement.

The numerical results presented in the paper are obtained based on the experimental and {{readily accessible}} parameters given in \cite{A8,A9,C4,D3,D9}:
$\omega_{m_{(1,2)}}/2\pi=10$GHz, 
$\lambda_L=1064$nm (optical wavelength),
$\omega_{b_{(1,2)}}/2\pi=40$MHz,
$\Phi=\pi/3$,
$\kappa_c/2\pi=2$MHz, 
$\kappa_{m_{(1,2)}}/2\pi=1$MHz,
$G_{_0}/2\pi=3$MHz,
$G_{m_1}/2\pi=2$MHz,
$G_{m_2}=G_{m_1}/2$,
$\gamma_{b_{(1,2)}}/2\pi=10^2$Hz, and the temperature {{$T = 10$ mK}}.
The OM coupling $G_{_0}$ can be achieved with a laser power $\mathcal{P}_L\simeq6.67$mW for $g_{_0}/2\pi=1$kHz at detuning $\tilde{\Delta}_{c}=\omega_{b}$ and $\Phi=\pi/3$ \cite{E8}. $G_{m_1}$ ($G_{m_2}$) corresponds to a microwave drive power $\mathcal{P}_{m_1}(\mathcal{P}_{m_2})\simeq0.91(0.46)$mW (the drive magnetic field $H_{d_1}(H_{d_2})\simeq8.7\times10^{-4}$T$(4.4\times10^{-4}$T)) with $g_{_m}/2\pi=20$Hz \cite{E9}.
Note that the much smaller micron-sized YIG crystal results in a much stronger yet dispersive magnon-phonon coupling, which is {{significantly greater than the $g_{_m}\sim10$ mHz}} used for a large YIG sphere \cite{A9,B1}.
All results in this article are obtained when the system is stable, as per the Routh-Hurwitz criterion \cite{F1}.\\

\begin{figure}[b]	
	\begin{center}
		\subfigure{\label{A6}\includegraphics[scale=0.45]{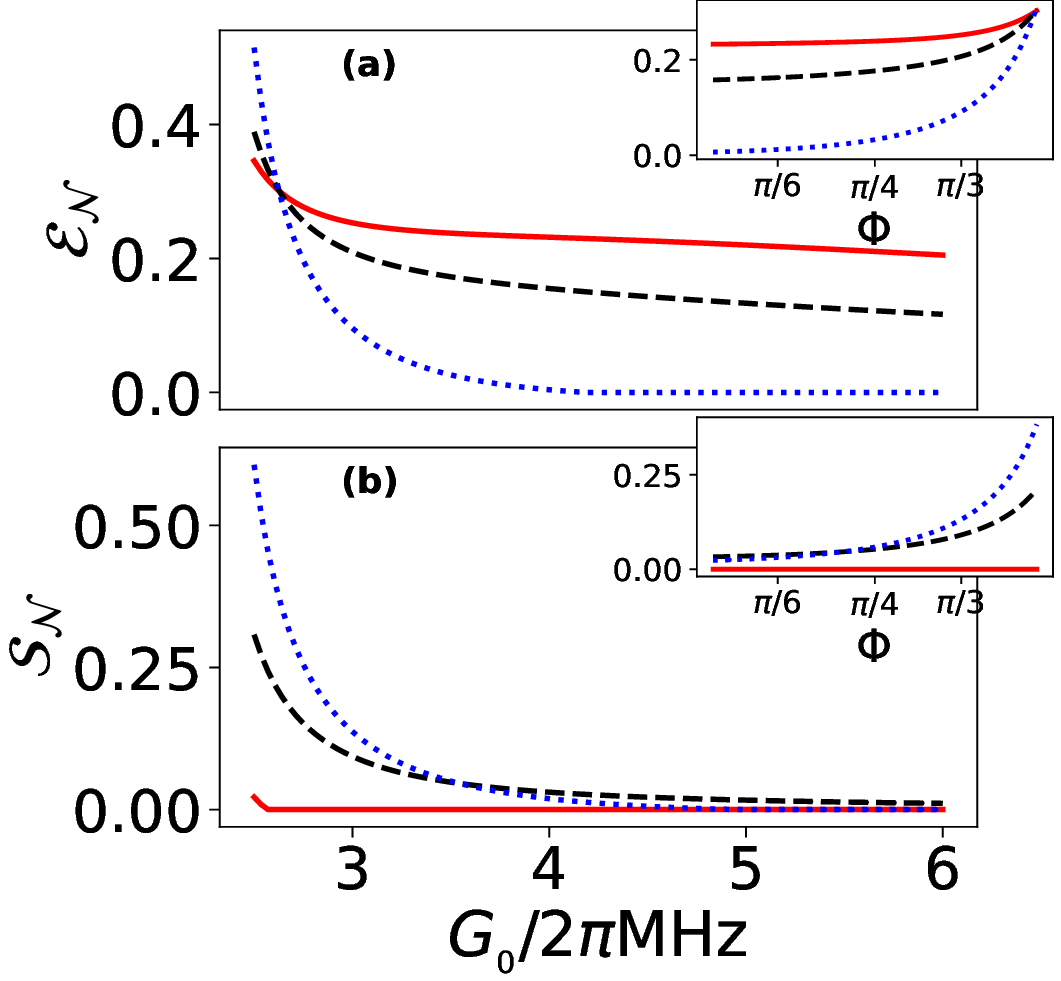}}
	\end{center}
   \captionsetup{justification=raggedright, singlelinecheck=false}
	\caption{{\small Optical cavity-magnon$_1$ (solid), optical cavity- magnon$_2$ (dashed), and magnon$_1$-magnon$_2$ (doted) (a) entanglement and (b) steering versus OM coupling $G_{_0}$ and $\Phi$ (see the inset, with $\tilde{G}_{_0}/2\pi=4$MHz). In the inset, we have kept the same optimized parameters. The parameters are the same as in Fig.\ref{fig:2}.}}
	\label{fig:3}
\end{figure}

In Fig.\ref{fig:2}, we present the study of quantum entanglement and {{asymmetric steering among the different optomagnonic subsystems as a function of key parameters such as detuning and temperature}}.  
Fig.\ref{fig:2}(a), (b) and (c) confirm, as previously mentioned, that the optimal condition to obtain entanglement corresponds to $\tilde{\Delta}_c=-\tilde{\Delta}_{m_{(1,2)}}\simeq\omega_{b}$.  
Notably, we {{observe that optomagnonic entanglement reaches a maximum value of 0.25 between the cavity and magnon$_1$, although no steering occurs in this case}}.  
Furthermore, the magnons' entanglement reaches a value of $0.1$ and shows a complex dependence {{in which the peak widths vary with the detuning frequency of the magnons}}, with strong steering compared to the other cases of optomagnonic steering of $0.14$.  
Moreover, temperature has a significant influence on the generation of quantum entanglement and asymmetric steering (Fig.\ref{fig:2}(d,h)).  
As the temperature increases, {{the entanglement gradually degrades and persists up to specific thresholds}} (i.e., robust against temperature):  
$\sim$($160$mK-$180$mK) for ($\mathcal{E}_{_{m_1-m_2}}-\mathcal{S}_{_{m_1-m_2}}$), $\sim$($240$mK-$160$mK) for $\mathcal{E}_{_{c-m_2}}-\mathcal{S}_{_{c-m_2}}$, and $\sim$$520$mK for $\mathcal{E}_{_{c-m_1}}$ with no asymmetric steering.\\

Note that a stronger OM coupling strength $G_{_0} > G_{_m}$ ensures the stability of the system. In this configuration, the anti-Stokes OM process by phonon absorption dominates the MM Stokes process by phonon emission.  
Fig.\ref{fig:3} shows the evolution of the entanglement and the asymmetric steering versus the OM coupling $G_{_0}$ and the angle between the incident light and the light reflected {{by the surfaces of the YIG microbridge}} $\Phi$ in the inset.  
However, the analysis reveals that an increase in OM coupling leads to a general decrease in entanglement. But in the specific case of entanglement between magnons, although it also decreases, it remains up to $G_{_0}/2\pi\sim 4.2$MHz and {{vanishes}} (i.e., it requires laser power $\mathcal{P}_L<13.09$mW for $g_{_0}/2\pi=1$kHz at detuning $\tilde{\Delta}_{c}=\omega_{b}$ and $\Phi=\pi/3$). Nevertheless, entanglement increases with increasing $\Phi$.  
Furthermore, in Fig.\ref{fig:3}(b), we observe that the asymmetric steering can be {{observed}} with a weak pulse field with OM coupling $G_0/2\pi<\sim2.56$MHz, {{which corresponds to a drive power}} $\mathcal{P}_L<2.73$mW.
\\

\begin{figure}[H]	
	\begin{center}
		\subfigure{\label{A4}\includegraphics[scale=0.37]{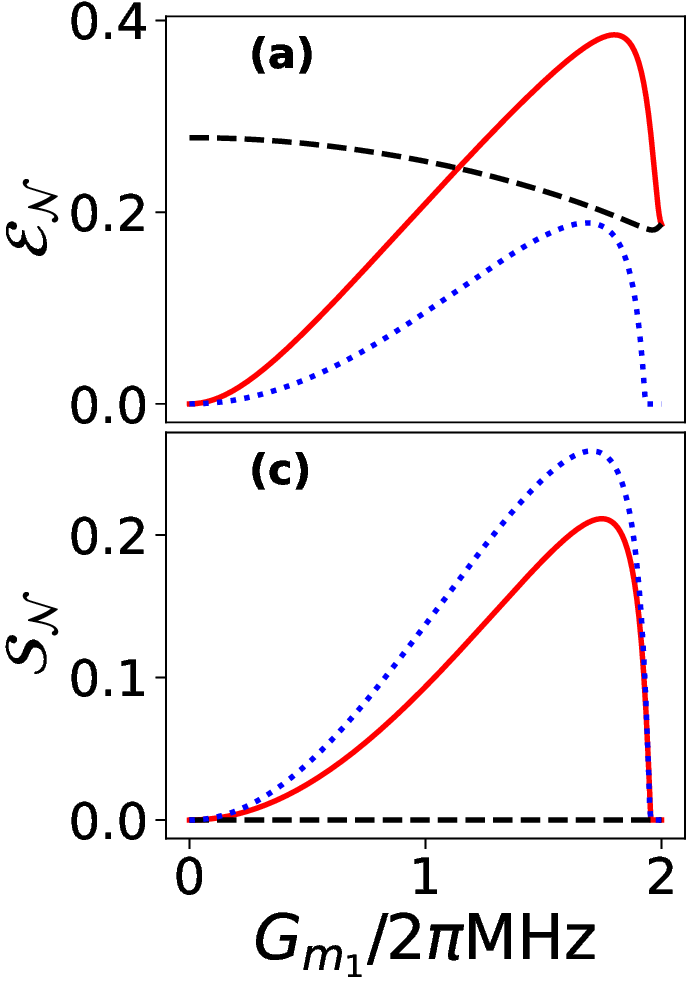}}
		\subfigure{\label{A6}\includegraphics[scale=0.37]{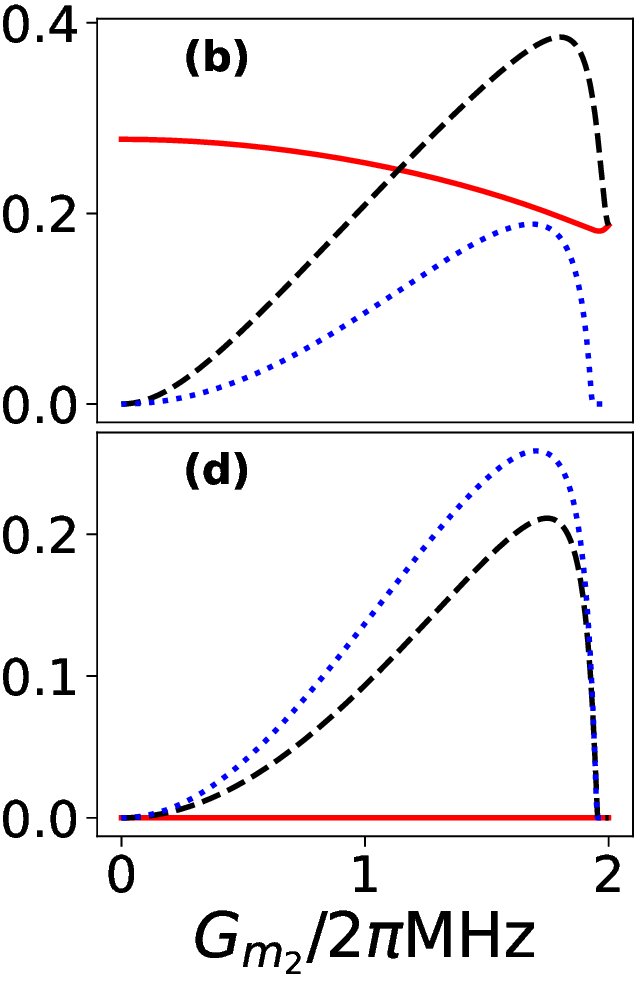}}
	\end{center}
	\caption{{\small Optical cavity-magnon$_1$ (solid), optical cavity-magnon$_2$ (dashed), and magnon$_1$-magnon$_2$ (doted) (a,b) entanglement and (c,d) steering versus (a,c) MM coupling of magnon$_1$ and (b,d) MM coupling of magnon$_2$. The parameters are the same as in Fig.\ref{fig:2}.}}
	\label{fig:4}
\end{figure}

Fig.\ref{fig:4}(a) and (b) show the evolution of entanglement versus the MM coupling $G_{_m}$ for the two magnons. {{In Fig.\ref{fig:4}(a) and (b), we maintain}} the MM coupling parameter of one magnon at $2\pi\times2$MHz while varying that of the other magnon. The results reveal a significant influence of the MM coupling parameter on the system: strong cavity-magnon entanglement is observed when the MM coupling value reaches $2\pi\times1.8$MHz. It should be noted that the magnon whose coupling is fixed at $2\pi\times2$MHz maintains a stronger entanglement with the cavity than the magnon whose coupling is variable.  
Furthermore, we observe that when the driving field becomes sufficiently strong, the MM coupling exceeds the weak-coupling regime ($G_{_m}\ll\omega_{b}$), allowing the rotating wave (RW) approximation for the cooling interaction $\propto\delta m^\dagger\delta d+\delta m\delta d^\dagger$. As a result, the counter-RW terms $\propto\delta m\delta d+\delta m^\dagger\delta d^\dagger$ of the linearised MM interaction become significant, activating the PDC and inducing MM entanglement. This entanglement is then transferred to the cavity-magnon and cavity-phonon subsystems when the cavity is in Stokes sideband resonance.  
However, as far as optomagnetic steering is concerned, the dependence of asymmetric steering on MM coupling reveals that no asymmetric steering occurs for cavity-magnon systems when one MM coupling parameter is fixed and the other is varied, and that optomagnonic asymmetric steering is generally absent when $G_{m_j}/2\pi=2$MHz is reached.\\

\begin{figure}[H]	
	\begin{center}		
		\subfigure{\label{A4}\includegraphics[scale=0.34]{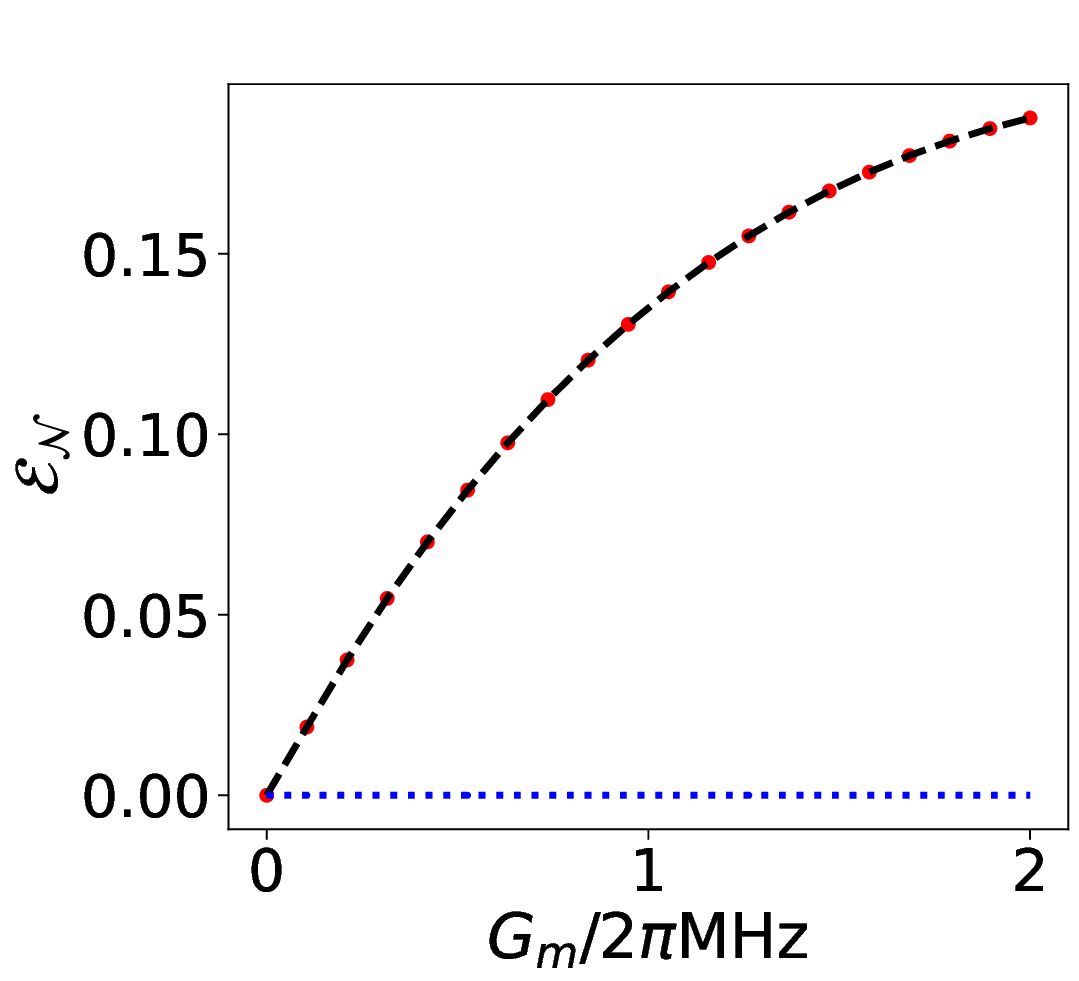}}
	\end{center}
	\caption{{\small Optical cavity-magnon$_1$ ($\bullet$ marker), optical cavity-magnon$_2$ (dashed), and magnon$_1$-magnon$_2$ (doted) entanglement versus MM coupling where $G_{_{m_1}}=G_{_{m_2}}=G_{_{m}}$. The parameters are the same as in Fig.\ref{fig:2}.}}
	\label{fig:5}
\end{figure}

However, a crucial observation concerning our proposed model is worth highlighting, namely, that optomagnonic asymmetric steering does not appear in the case where $G_{m_1}=G_{m_2}$. The entanglement between cavity-magnon$_1$ and {{cavity-magnon$_2$ becomes equal}}. More importantly, however, entanglement between magnons becomes impossible when the magnon-magnon couplings are equal (Fig.\ref{fig:5}).\\

\begin{figure}[t]	
	\begin{center}
		\subfigure{\label{A4}\includegraphics[scale=0.36]{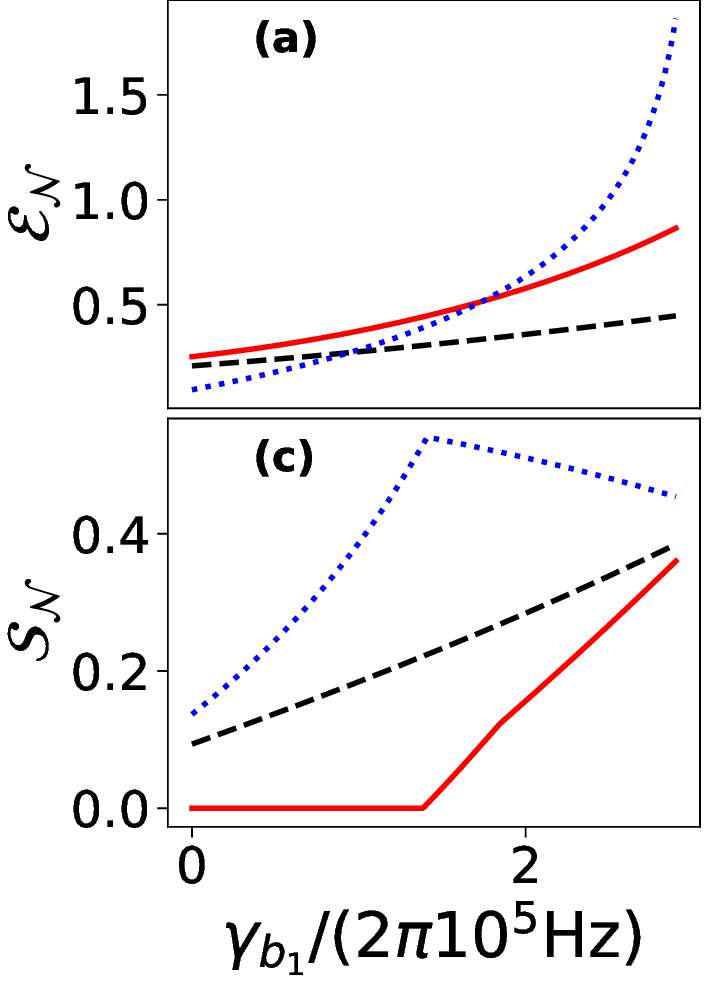}}
		\subfigure{\label{A4}\includegraphics[scale=0.36]{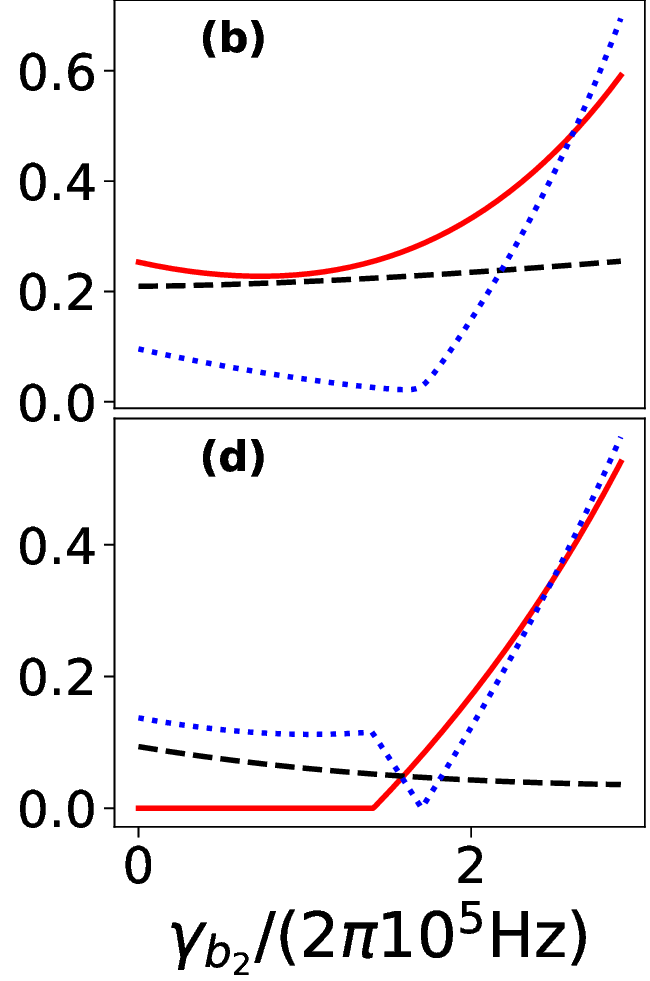}}
	\end{center}
   \captionsetup{justification=raggedright, singlelinecheck=false}
	\caption{{\small Optical cavity-magnon$_1$ (solid), optical cavity- magnon$_2$ (dashed), and magnon$_1$-magnon$_2$ (doted) entanglement-steering versus the damping rate of (a-c) phonon$_1$ mode, and (b-d) of the damping rate of phonon$_2$ mode. The parameters are the same as in Fig.\ref{fig:2}.}}
	\label{fig:6}
\end{figure}

The additional mirror introduces additional mechanical damping of the vibrational mode, but its effect on entanglement remains negligible. Fig.\ref{fig:6} shows the influence of the mechanical damping of the phonons on entanglement. The entanglement remains robust as the damping of the phonon$_1$ increases. On the other hand, when the damping of the phonon$_2$ is varied, the entanglement decreases {{as}} the magnon-magnon entanglement increases up to $\gamma_{b_2}/2\pi\simeq1.6\times10^5$Hz (Quality factor $\mathcal{Q}_2=250$), and then increases again. Moreover, the asymmetric steering versus phonon$_1$ damping rate exhibits significant behavior. For the cavity-magnon$_1$ steering, a significant effect is observed at {{the value}} $\gamma_{b_1}/2\pi\simeq1.38\times10^5$Hz ($\mathcal{Q}_1=\mathcal{Q}_2/1380$). For the magnon1-magnon2 steering, it initially increases and then decreases at this critical point.  
However, for the asymmetric steering versus phonon$_2$ damping rate, the cavity-magnon$_1$ steering behaves similarly, appearing at {{the value}} $\gamma_{b_2}/2\pi\simeq1.41\times10^5$Hz ($\mathcal{Q}_2=283$). As for the magnon$_1$-magnon$_2$ steering, it initially decreases up to the point where the quality factor reaches $\mathcal{Q}_2=235$, {{vanishing at this point, and then increasing again}} (Fig.\ref{fig:6}(b-d)).\\

Lastly, to ensure the validity of the approximations that were made in our model with the results of all the previous figures, we assume that the number of magnon excitations $\braket{m_1^\dagger m_1}\ll2N_0s=5N_0$, where $s=\frac{5}{2}$ is the spin number of the ground state Fe$^{3+}$ ion in YIG. For a YIG bridge structure of a given geometry, the number of spins $N_0=4.22\times 10^{10}$. $|\braket{m_1}| \simeq 6.77 \times 10^4$ corresponds to the MM coupling $G_{m_1}/2\pi=2$MHz and the Rabi frequency $\Omega_1\simeq1.7\times10^{13}$Hz, which leads to $|\braket{m_1^\dagger m_1}| \simeq 4.57 \times 10^9\ll5N_0=2.2\times10^{11}$. This condition is, therefore, satisfied. Concerning the second MM part ($m_2$), given also that $|\braket{m_2}|\ll5N_0$, since it can be seen by using $|\braket{m_2}|\ll|\braket{m_1}|$. We used two identical YIG microbridges, providing a simplified framework for our analysis as our numerical study incorporates phonon modes with identical frequencies of $\omega_{b}/2\pi=40$MHz.

Furthermore, a high pumping intensity is adopted for the magnon modes, which can introduce unwanted non-linearities due to the non-linear Kerr term $\mathcal{K}m_j^{\dagger}m_jm_j^{\dagger}m_j$ in the Hamiltonian. To ensure that the Kerr effect $\mathcal{K}$ remains negligible, the condition $\mathcal{K}|\braket{m_j}|^3\ll\Omega_{j}$ must be satisfied, thus guaranteeing the validity of our linearized model. The positive Kerr coefficient $\mathcal{K}$ is inversely proportional to the sample volume and is given in \cite{X0} by $\mathcal{K}=(\mathcal{K}_{\text{an}}\mu_0\gamma^2)/(\mathcal{M}V^2)$, where $\mathcal{K}_{\text{an}}$ is the first-order anisotropy constant, $\mathcal{M}$ is the saturation magnetization, and the other parameters are already given in the paper. However, for a $1$mm diameter YIG sphere, $\mathcal{K}/2\pi = 0.1$nHz \cite{X0,X1}. For a YIG sphere of diameter $250\mu$m, $\mathcal{K}/2\pi=0.1\times4^3=6.4$nHz \cite{B1,B02,B03}. According to the parameters considered in our study, the condition $\mathcal{K}\ll\Omega_{1}/|\bar{m_1}|^3$ must be strictly satisfied, which in our case implies $\mathcal{K}\ll\Omega_{1}/|\braket{m_1}|^3\simeq1.7\times10^{13}/(3.1\times10^{14})\sim2\pi\times8.57$mHz. We have chosen ${m_1}$ (as $|\braket{m_2}|\ll|\braket{m_1}|$) to refer to the minimal case.

\section{Conclusion}
\label{sec6}

In this paper, we have proposed a novel ring-optomagnomechanical configuration to generate entanglement in a cavity system composed of two magnomechanical systems. In this configuration, each phonon mode is coupled to a magnon mode in YIG microbridges, while being simultaneously coupled to the single optical cavity via magnetostriction and radiation-pressure forces. Importantly, we showed that the proposed system can be adopted to prepare a stationary entangled state between magnons and optical photons by exploiting the dispersive interactions of magnetostriction and radiation pressure. Moreover, we studied the case when the cavity is excited using a red-detuned microwave field, while the YIG microbridges {{are essentially excited using a blue-detuned field}}. Accordingly, we demonstrated that the magnomechanical parametric down-conversion establishes magnon-phonon entanglement. However, the optomechanical beam-splitter interaction {{enabled photon-phonon state swapping}}. Hence, the phonon modes act as intermediaries, creating an entanglement with the common photons, resulting in a magnon-magnon macroscopic entangled state \cite{ZI}.\\

Our results, based on experimentally achievable parameters, showed that the entanglement can achieve important bounds {{with respect to temperature robustness}}. Nevertheless, we observed that the magnon$_1$-magnon$_2$ entanglement disappears when the magnomechanical couplings are identical. Hence, these results open up new perspectives for hybrid quantum devices that can {{enable envisioning}} a promising experimental realization. In fact, this innovative optomagnetic-ring system could open up new {{applications in}} magnonics-based quantum information processing, \cite{F2,F3,F4} and quantum networks \cite{C1}.


\section*{Appendix A}
\label{appendix:A}
\renewcommand{\theequation}{A\arabic{equation}}
\setcounter{equation}{0} 

In this appendix, we give a detailed derivation of the total Hamiltonian, to obtain the drift matrix of Eq.\ref{6}, using the QLE of Eq.\ref{2}.

Once we have obtained the interaction picture of the total Hamiltonian, using a unitary transformation $\mathbf{U}=\exp [i(\omega_Lc^\dagger c+\sum_{j=1,2}\omega_0m_j^\dagger m_{j})t]$, we assign a change to the operator as follows $m_j e^{i\omega_0t}\to m_j$ and $c e^{i\omega_Lt}\to c$. It is given as follows
\begin{eqnarray}
	\label{X1}
	\tilde{\mathbf{H}}&=&\mathbf{U}\mathbf{H}\mathbf{U}^\dagger-\text{i}\mathbf{U}\partial_t \mathbf{U}^\dagger\nonumber\\
	&=&(\omega_c-\omega_L) c^\dagger c +\sum_{j=1,2}(\omega_{m_j}-\omega_0) m_j^\dagger m_j+\frac{\omega_{b_j}}{2}(q_{_j}^2+p_{_j}^2) \nonumber\\
	&+& \bar{g}_{0j}c^\dagger c q_{_j}+g_{m_j}m_j^\dagger m_jq_{_j}+i\big[\big(Ec^\dagger+\Omega_j m_j^\dagger-\text{H.C}\big].
\end{eqnarray}

Therefore, the quantum dynamics of the system considered by adding dissipative noise and input noise for each mode can be governed by the following QLEs.  We obtain a set of QLEs for the system as follows
\begin{eqnarray}
	\label{X2}
		\dot{c}&=& -i(\omega_c-\omega_L)c-{\kappa}_{c}c-i\sum_{j=1,2}i\bar{g}_{0}cq_{_j}+E + \sqrt{2\kappa_{c}} c^{in},\nonumber\\
	\dot{m}_j &=& -i(\omega_{m_j}-\omega_0)m_j-\kappa_{m_j}m_j-ig_{m_j}m_jq_j+\Omega_j\nonumber\\
	&+&\sqrt{2\kappa_{m_j}} m^{in}_j,\nonumber\\
	\dot{q}_j&=&\omega_{b_j}p_j,\nonumber\\
	\dot{p}_j&=&-\omega_{b_j}q_j-g_{_{mj}}m_j^\dagger m_j-\sum_{j=1,2}\bar{g}_{0j}c^\dagger c-\gamma_{b_j} p_j+\xi_j,
\end{eqnarray}
where the operators are defined in the main text.\\

In the following and as mentioned earlier, we linearise the QLEs of Eq.\ref{X2} in the form in the main text of Eq.\ref{4}, around the steady state value and their quantum fluctuation. However, we have already given a series of algebraic equations for the steady-state values in Eq.\ref{4}.

By neglecting the second-order fluctuation terms, as the mean value of the physical quantity is much bigger than its fluctuation, we obtain linearized QLEs for the quantum fluctuations as follows
{\normalsize 
\begin{eqnarray}
	\label{X3}
	\dot{\tilde{c}}&=& -i\tilde{\Delta}_c\tilde{c}-{\kappa}_{c}\tilde{c}-i\sum_{j=1,2}\bar{g}_{0}\alpha_{s}\tilde{q}_{_j} +\sqrt{2\kappa_{c}} \tilde{c}^{in},\nonumber\\
	\dot{\tilde{m}}_j &=& -i\tilde{\Delta}_{m_j}m_j-\kappa_{m_j}m_j-ig_{m_j}m_{js}\tilde{q}_j+\sqrt{2\kappa_{m_j}} \tilde{m}^{in}_j,\nonumber\\
	\dot{\tilde{q}}_j&=&\omega_{b_j}\tilde{p}_j,\nonumber\\
	\dot{\tilde{p}}_j&=&-\omega_{b_j}\tilde{q}_j+\frac{\bar{G}_{_0}}{i\sqrt{2}}(\tilde{c}^\dagger- \tilde{c})-\frac{G_{m_{j}}}{i\sqrt{2}}(\tilde{m}_j^\dagger- \tilde{m}_j)\nonumber\\
	&-&\gamma_{b_j} \tilde{p}_j+\xi_j,
\end{eqnarray}	
}
where $\tilde{\Delta}_a=\omega_a-\omega_L -\sum_{j=1,2}(-1)^{{j}}{g}_{_0}\cos^2({\Phi}/{2})q_{js}$, $\tilde{\Delta}_{m_j}=\omega_m-\omega_0+g_{m_j}q_{js}$, and $\bar{G}_{_0}=(-1)^{{j}}G_{_0}$ where ${G}_{_0}=i{g}_{_0}\sqrt{2}\cos^2({\Phi}/{2})\alpha_{s}$.
Meanwhile, the introduction of a set of quadrature fluctuations and input noise operators
\begin{equation}
	\label{X4}
	\begin{aligned}
		\tilde{I}_c&=\frac{\tilde{c} +\tilde{c}^\dagger}{\sqrt{2}}~~&,&~~ \tilde{J}_c&=&\frac{\tilde{c}-\tilde{c}^\dagger}{i\sqrt{2}}\\
		\tilde{I}_{m_j}&=\frac{ \tilde{m_j} +\tilde{m_j}^\dagger}{\sqrt{2}}~~&,&~~ \tilde{J}_{m_j}&=&\frac{\tilde{m_j}-\tilde{m_j}^\dagger}{i\sqrt{2}}.\\
	\end{aligned}
\end{equation}

Eq.\ref{X4} can be rewritten as
{\normalsize { \begin{eqnarray}
	\label{X5}
	\dot{\tilde{I}}_c&=&\tilde{\Delta}_c \tilde{J}_c-{\kappa}_c \tilde{I}_c +\bar{G}_{_{\sum}}\tilde{q}_j+\sqrt{2\kappa_{_c}} \tilde{I}_c^{in},\nonumber\\
	\dot{\tilde{J}}_c&=&-\tilde{\Delta}_c\tilde{I}_c-{\kappa}_a\tilde{J}_c+\sqrt{2\kappa_{_c}} \tilde{J}_c^{in},\nonumber\\
	\dot{\tilde{I}}_{m_j}&=&\tilde{\Delta}_{m_j} \tilde{J}_{m_j}-{\kappa}_{m_j}\tilde{I}_{m_j} -{G}_{m_{j}}\tilde{q}_j+\sqrt{2\kappa_{{m_j}}} \tilde{I}_{m_j}^{in},\nonumber\\
	\dot{\tilde{J}}_{m_j}&=&-\tilde{\Delta}_{m_j}\tilde{I}_{m_j}-{\kappa}_{m_j}\tilde{J}_{m_j}+\sqrt{2\kappa_{m_j}} \tilde{J}_{m_j}^{in},\nonumber\\
	\dot{\tilde{q}}_j&=&\omega_{b_1}\delta p_1,\nonumber\\	
	\dot{\tilde{p}}_j&=&-\omega_{b_1}\tilde{q}_j-{\bar{G}_{_{0}}}\tilde{J}_a+{G_{m_{j}}}\tilde{J}_{m_j}-\gamma_{b_j}\delta p_j+\xi_1,
\end{eqnarray}}}
where $\bar{G}_{\sum}=\sum_j\bar{G}_{_0}=\sum_j(-1)^{{j}}G_{_0}$, and 
\begin{equation}
	\label{X6}
	\begin{aligned}
		\tilde{I}_c^{in}&=\frac{\tilde{c}^{in} +\tilde{c}^{\dagger in}}{\sqrt{2}}~~&,&~~ \tilde{J}_c^{in}&=&\frac{\tilde{c}^{in}-\tilde{c}^{\dagger in}}{i\sqrt{2}}\\
		\tilde{I}_{m_j}^{in}&=\frac{ \tilde{m_j}^{in} +\tilde{m_j}^{\dagger in}}{\sqrt{2}}~~&,&~~ \tilde{J}_{m_j}^{in}&=&\frac{\tilde{m_j}^{in}-\tilde{m_j}^{\dagger in}}{i\sqrt{2}}.
	\end{aligned}
\end{equation}
Other parameters are defined in the main text.
The Eq.\ref{X6} can be given in a compact form as
\begin{equation}
	\label{X7}
	\dot{\Xi}(t)=\Upsilon\Xi(t)+\mathcal{N}(t),
\end{equation}
where 
\begin{equation}
	\label{X8}
\Xi(t)= [\sqrt{2{\kappa}_{c}}\tilde{\Theta}_{c},\Xi_{m_1},\Xi_{m_2}]^T
\end{equation}
 with $\Xi_{m_j}= \sqrt{2{\kappa}_{m_j}}\tilde{\Theta}_{m_j}\oplus[\tilde{q}_j,\tilde{p}_j]$
and  $\tilde{\Theta}_\varphi=[\tilde{I}_\varphi,\tilde{J}_\varphi]$, with ($\varphi=c,m_j$).\\

$\mathcal{N}(t)=[\sqrt{2{\kappa}_{c}}\tilde{\Theta}^{in}_{c},\mathcal{N}_{m_1},\mathcal{N}_{m_2}]^T$is the noise vector,
where $\tilde{\Theta}^{in}_\varphi=[\tilde{I}^{in}_\varphi,\tilde{J}^{in}_\varphi]$, and 
$\mathcal{N}_{m_j}=[\sqrt{2{\kappa}_{m_j}}\tilde{\Theta}^{in}_{m_j}\oplus[0,\xi_j]]$, using the quadrature of the input noise vector defined in Eq.\ref{X6}. $\Upsilon$ is the drift matrix defined and takes the form in Eq.\ref{6}.


\begin{thebibliography}{9}
\bibitem{D7}  However, utilizing a thin film in the optomagnomechanical system is not feasible, as magnons and high-frequency phonons can couple and create a strong dispersive linear coupling. The non-linearity of the magnon-phonon interaction is necessary to probe the non-local correlation.
\bibitem{E8}  The optomechanical coupling $g_{_0}$ is larger than the magnomechanical coupling $g_{_m}$, being around $\sim0.1-1$kHz for a micron-sized mirror as used in \cite{D5,D6}.
\bibitem{E9} The driving magnetic field is defined as $H_d=\sqrt{2\mu_0\mathcal{P}_m/(LWc)}$, where $\mu_0$ and $c$ are the magnetic permeability and velocity of electromagnetic waves in vacuum, $L$ and $W$ are the length and width of the YIG microbridge cross-section, respectively. In addition, $\mathcal{P}_m= (H_d^2/2\mu_0)LWc$ is the microwave drive power, where $H_d^2/2\mu_0$ is the time-averaged energy density per unit volume.
\bibitem{ZI} The magnon mode represents a collective mode that contains a large number of spins ($N_0=4.22\times 10^{10}$ for a $5\times2 \times 1 \mu m^3$ YIG-microbridge). We consider that it is on a macroscopic scale and can be described as a macroscopic quantum state.
\end{thebibliography}
\end{document}